\documentclass[sigconf]{acmart}

\AtBeginDocument{%
  \providecommand\BibTeX{{%
    \normalfont B\kern-0.5em{\scshape i\kern-0.25em b}\kern-0.8em\TeX}}}

\usepackage{amsmath}
\usepackage{graphicx,nicefrac}
\usepackage{caption}
\usepackage{subcaption}
\usepackage{cleveref}
\usepackage{adjustbox}
\usepackage{multirow}

\usepackage{hyperref}
\hypersetup{
    colorlinks=true,
    linkcolor=blue,
    filecolor=magenta,      
    urlcolor=cyan,
}

\usepackage{amsthm}

\newtheorem{theorem}{Theorem}[section]
\newtheorem{definition}[theorem]{Definition}

\acmConference[AdvML '21]{AdvML '21: Workshop on Adversarial Learning Methods for Machine Learning and Data Mining}{August 15, 2021}{Singapore (virtual)}
\acmBooktitle{AdvML '21: Workshop on Adversarial Learning Methods for Machine Learning and Data Mining, August 15, 2021, Singapore (virtual)}
\usepackage[skip=2pt]{caption}

\begin{document}


\title{Understanding the Effects of Adversarial Personalized Ranking Optimization Method on Recommendation Quality}




\author{Vito Walter Anelli, Yashar Deldjoo, Tommaso Di Noia,  Felice Antonio Merra}
\affiliation{%
  \country{Polytechnic University of Bari, Italy}
}
\email{name.surname@poliba.it}

\authornote{The authors are in alphabetical order. Corresponding author: Felice Antonio Merra (\texttt{felice.merra@poliba.it}).}

\begin{abstract}
Recommender systems (RSs) employ user-item feedback, e.g., ratings, to match customers to personalized lists of products. Approaches to top-k recommendation mainly rely on Learning-To-Rank algorithms and, among them, the most widely adopted is  Bayesian Personalized Ranking (BPR), which bases on a pair-wise optimization approach. Recently, BPR has been found vulnerable against adversarial perturbations of its model parameters. Adversarial Personalized Ranking (APR) mitigates this issue by robustifying BPR via an adversarial training procedure. The empirical improvements of APR's accuracy performance on BPR have led to its wide use in several recommender models. However, a key overlooked aspect has been the beyond-accuracy performance of APR, i.e., novelty, coverage, and amplification of popularity bias, considering that recent results suggest that BPR, the building block of APR, is sensitive to the intensification of biases and reduction of recommendation novelty. In this work, we model the learning characteristics of the BPR and APR optimization frameworks to give mathematical evidence that, when the feedback data have a tailed distribution, APR amplifies the popularity bias more than BPR due to an unbalanced number of received positive updates from short-head items. Using matrix factorization (MF), we empirically validate the theoretical results by performing preliminary experiments on two public datasets to compare BPR-MF and APR-MF performance on accuracy and beyond-accuracy metrics. The experimental results consistently show the degradation of novelty and coverage measures and a worrying amplification of bias.

\end{abstract}


\maketitle

\section{Introduction and Motivation}\label{sec:introduction}
Machine-learned models such as latent factor models (LFMs) have significantly advanced the capability of recommender systems (RSs) to be faster and more accurate. To make recommendations, modern RSs often employ Bayesian Personalized Ranking (BPR)~\cite{DBLP:conf/uai/RendleFGS09}, a pairwise ranking optimization framework that uses item pairs as training data and optimizes it for correctly ranking item pairs. BPR is currently a state-of-the-art optimization framework adopted in many research works~\cite{DBLP:journals/corr/HidasiKBT15, DBLP:conf/sigir/ChenZ0NLC17, DBLP:conf/sigir/Wang0WFC19, DBLP:conf/ecir/AnelliDNFN21}. 

The recent survey by~\citet{adversarial2021survey} has shown that recommender models are fragile against \textit{adversarial attacks}, i.e., small but non-random perturbations added to the model data to cause recommendation performance (i.e., parameters~\cite{anelli2020multistep,DBLP:conf/sigir/0001HDC18}, content~\cite{DBLP:conf/dsn/NoiaMM20,DBLP:conf/www/LiuL21,DBLP:conf/wsdm/CohenSJA21,anelli2020empirical}, user-item matrix~\cite{DBLP:conf/esws/AnelliDNSM20,DBLP:conf/sigir/DeldjooNSM20}). Several works have shown the vulnerability of LFMs trained with BPR under adversarial attacks, for instance,~\citet{DBLP:conf/sigir/0001HDC18} empirically verify that adversarial perturbation of BPR-MF, i.e., a matrix factorization (MF) model trained with BPR, decreases the nDCG metric value by -26.3\%. For example,~\citet{DBLP:conf/sigir/YuanYB19} show the same degradation on collaborative auto-encoder (CAE) models and~\citet{DBLP:journals/tkde/TangDHYTC20} validate it on visual-based recommenders.

To address this issue, as a defensive remedy,~\citet{DBLP:conf/sigir/0001HDC18} propose Adversarial Personalized Ranking (APR), a novel optimization strategy to robustify BPR against adversarial perturbations. Based on the \textit{adversarial training} procedure proposed by~\citet{DBLP:journals/corr/GoodfellowSS14}, APR extends BPR by integrating the BPR-objective function with an additional regularization term, named \textit{adversarial regularizer}, that quantifies the loss value when the model parameters are adversarially perturbed. The robustified version of BPR showed a nDCG reduction of only -2.9\% on MF~\cite{DBLP:conf/sigir/0001HDC18},  a protection effect confirmed also on other models such as CAE~\cite{DBLP:conf/sigir/YuanYB19}, TF~\cite{DBLP:conf/recsys/ChenL19}, and VBPR~\cite{DBLP:journals/tkde/TangDHYTC20}. The key insight is that APR not only improves the defensive capability of RS (robustness under adversarial attacks) but also their generalization performance in normal item recommendation tasks. For instance,~\citet{DBLP:conf/sigir/0001HDC18} show that for optimizing MF, if APR is used instead of BPR, a relative improvement of +11\% on accuracy performance is achieved when compared to BPR results. 

Given the gained performances obtained in both robustness and accuracy dimensions, we have recently witnessed the application of APR in a growing number of research works. More than 15 articles present novel recommendation algorithms incorporating the APR as the core optimization framework~\cite{DBLP:conf/sigir/0001HDC18,DBLP:conf/ijcnn/YuanYB19,DBLP:conf/sigir/YuanYB19,DBLP:conf/sigir/TranSL19,DBLP:conf/recsys/ChenL19,DBLP:conf/www/ParkC19,DBLP:conf/www/DaiSZLW19,8924766,https://doi.org/10.1049/cje.2020.05.004,9338819,DBLP:conf/wsdm/LiW020,DBLP:conf/cikm/YuanYB20,DBLP:journals/access/WangH20a,DBLP:journals/tkde/TangDHYTC20,Hu2021}. These examples underline the popularity of the adversarial ranking-based procedure, i.e., APR, for various item recommendation tasks. However, given the sensitivity of BPR against popularity bias reported in recent works~\cite{DBLP:journals/umuai/JannachLKJ15, DBLP:conf/recsys/AbdollahpouriBM17,DBLP:conf/sigir/ZhuWC20,DBLP:journals/ipm/BorattoFM21}, the question remains as to how much APR is vulnerable against the amplification of popularity bias considering that BPR is the APR building block. 


Motivated by this observation, the main contributions of this work include: \textbf{(1)}  the presentation of a formal analysis to identify whether APR is affected by popularity amplification bias, and highlighting how difference such bias is in comparison with BPR (the core building block used in APR); \textbf{(2)} the empirical verification of the existence of a trade-off between accuracy and beyond-accuracy measures and popularity bias in APR --- thus there is no free cake!  An experimental evaluation has been carried out on two recommendation datasets using MF as the base ML model. The results motivate the design of novel pairwise robust learning procedures that can strike a more meaningful balance between accuracy, beyond accuracy, and bias amplification.
\section{Formal Analysis}\label{sec:proposal}

\noindent\textbf{Preliminaries.} Let $F \subset U \times I$ be the matrix of user-item feedback, where $U$ is the set of $M$ users $\{u_1, u_2, ..., u_M\}$ and $I$ is the set of $N$ items $\{i_1, i_2, ..., i_N\}$. 
The item recommendation task builds a user's personalized list of $k$ items ranked by predicted relevance scores. Given a user $u \in U$, the rank of a not-interacted item $i \in I$ is defined via the bijective function in $I$ as $\hat{r}(i|u)$. The ranking function $\hat{r}(\cdot)$ is based on the predicted value of the preference score function $\hat{s}(\cdot|\Theta)$. $\Theta$ represents the ML recommender's model parameters, e.g., matrix factorization (MF)~\cite{DBLP:journals/computer/KorenBV09}. To build the top-$k$ recommendation list associated with $u$, the user's not-interacted items are sorted in decreasing order by the predicted score. Formally, the rank of each item is defined as $ \hat{r}(i \mid u):=\big\{ |\{j: \hat{s}(j \mid u) \geq \hat{s}(i \mid u)\}|, i,j \in I \backslash I^{+}_{u} \big\}$, where $ I^{+}_{u}$ is the list of (positive) items already seen by $u$.

\noindent\textbf{BPR.}
The recommender model parameters ($\Theta$) are learned with optimization procedures.  Bayesian personalized ranking (BPR) is a standard strategy in several RSs~\cite{DBLP:journals/corr/HidasiKBT15,DBLP:conf/sigir/ChenZ0NLC17,DBLP:conf/sigir/Wang0WFC19}. It assumes that given a user $u$, the score $\hat{s}(i|u)$ predicted on an already interacted item $i \in I^{+}_{u}$ should be higher than the one estimated for a not-interacted item $j \in I \backslash I^{+}_{u}$. Commonly, the first item is called \textit{positive}, while the seconds \textit{negative}. A user $u$, a positive item $i$, and a negative item $j$ form $(u,i,j)$ a training triplet. The full set of pair-wise preferences $D_F	\subseteq U \times I \times I$ is composed by all the triplets $(u,i,j)$ such that $(u,i,j) \in D_F : \iff	\Big( i \in I^{+}_{u} \land	 j \in I \backslash I^{+}_{u} \Big) $. BPR associates a negative item $j$ to each $(u, i)$-pair by uniformly sampling $j$ from the set of $u$ not-interacted ones ($I \backslash I^{+}_{u}$). Since BPR associates a single negative item to each recorded pair of interactions, it follows that the size of $D_F$ is equal to the number of recorded preferences, with $|D_F| \ll |F|$. Let $\sigma(\cdot)$ bet the sigmoid function, BPR learns $\Theta$ to optimize
\begin{equation}
  \underset{\Theta}{\operatorname{argmin}} - \sum_{(u, i, j) \in D_{F}} \ln \sigma(\hat{s}(i|u) - \hat{s}(j|u)) = \underset{\Theta}{\operatorname{argmin}} \mathcal{L}_{BPR}
  \label{eq:maxbpr}
\end{equation}
and, using the stochastic gradient descent (SGD), the model parameters are updated as $\Theta \leftarrow \Theta + \eta (1-\sigma(\hat{s}_{uij}(\Theta))) \frac{\partial \hat{s}_{uij}(\Theta)}{\partial \Theta}$, where $\eta$ is the learning rate. In the following, we will use $\hat{s}_{uij}(\Theta)$ to indicate the $\hat{s}(i|u)-\hat{s}(j|u)$ for lightening the formalism.

\noindent\textbf{APR.} We define the adversarial perturbation ($\Delta_{adv}$) as 
\begin{equation}
    \small
  \Delta_{adv} := \underset{\Delta, ||\Delta|| \leq \epsilon}{\operatorname{argmax}} \mathcal{L}_{BPR}(\hat{\Theta} + \Delta)
  \label{eq:adv_attacks}
\end{equation}
where $\epsilon$ is the perturbation budget to limit the maximum amount of noise added to the $\Theta$, $||\cdot||$ is the $L_2$-norm, and $\hat{\Theta}$ denotes the fixed model parameters on which the perturbation is evaluated. The intuition is that building a perturbation that increases the model's loss reduces the recommendation performance. Inspired by the fast gradient sign method by~\citet{DBLP:journals/corr/GoodfellowSS14}, ~\citet{DBLP:conf/sigir/0001HDC18} solved~\Cref{eq:adv_attacks} by linearizing the objective function $\mathcal{L}_{BPR}$ as $\Delta_{adv}= \epsilon \cdot \nicefrac{\Gamma}{\|\Gamma\|} $, where $\Gamma=\nicefrac{\partial \mathcal{L}_{BPR}(\hat{\Theta}+\Delta)}{\partial \Delta}$.

To robustify, and stabilize, the BPR-learned model against $\Delta_{adv}$, ~\citet{DBLP:conf/sigir/0001HDC18} proposed to use an \textit{adversarial training} procedure. The procedure, named adversarial personalized ranking (APR), learns $\Theta$ within a minimax optimization game
\begin{equation}
\arg \min _{\Theta} \max _{\Delta_{adv},\|\Delta_{adv}\| \leq \epsilon} \mathcal{L}_{BPR}(\Theta)+\alpha \mathcal{L}_{BPR}(\Theta+\Delta_{adv})
\label{eq:minimax}
\end{equation}
where the BPR loss ($\mathcal{L}_{BPR}$) and the regularization term ($\alpha\mathcal{L}_{BPR}(\Theta+\Delta_{adv})$) composed $\mathcal{L}_{APR}(\Theta)$, the APR objective function, where $\alpha$ is named \textit{adversarial regularization coefficient}. The additional regularization term, named \text{adversarial regularizer}, is the loss obtained when $\Delta_{adv}$ is added to $\Theta$ to \textbf{maximize} the model objective (see~\Cref{eq:adv_attacks}). It follows that, being $\Delta_{adv}$ fixed, APR \textbf{minimizes} both the standard BPR loss $\mathcal{L}_{BPR}$ with, and without, $\Delta_{adv}$. The aim of APR is to learn a model that is able to correctly distinguish the positive and negative items also in adversarial settings. As performed in~\cite{DBLP:conf/sigir/0001HDC18}, $\Theta$ updates are computed as follows $\Theta \leftarrow \Theta + \eta \big[ (1-\sigma(\hat{s}_{uij}(\Theta))) \frac{\partial \hat{s}_{uij}(\Theta)}{\partial \Theta} + \alpha (1-\sigma(\hat{s}_{uij}(\Theta + \Delta_{adv}))) \frac{\partial \hat{s}_{uij}(\Theta + \Delta_{adv})}{\partial \Theta} \big]$.

\subsection{Gradient Magnitudes}~\label{subsec_grad-gmanitudes}
We compare BPR and APR, studying their gradient magnitudes.

\subsubsection{Bayesian Gradient Magnitude}
The $\Theta$ updates with BPR depend on the learning rate $\eta$, the partial derivative of the difference of predicted scores $\hat{s}_{uij}(\Theta)$, and a multiplicative scalar $(1-\sigma(\hat{s}_{uij}(\Theta)))$. Following ~\citet{DBLP:conf/wsdm/RendleF14}, we define the \textit{Bayesian gradient magnitude} ($\omega$) on $(u,i,j)$ triplet as $  \omega_{uij} := (1-\sigma(\hat{s}_{uij}(\Theta)))$. This multiplicative scalar indicates how much the current model represented by $\Theta$ is performing in recognizing that $u$ prefers $i$ more than $j$. The update significantly changes $\Theta$ when $
  \omega_{uij} \simeq 1 \implies \big( \sigma(\hat{s}_{uij}(\Theta)) \simeq 0 \iff \hat{s}_{ui}(\Theta) \ll \hat{s}_{uj}(\Theta) \big)$.
In this circumstance, the preference score $\hat{s}_{uj}$ predicted for the negative item $j$ is bigger than the one predicted on the positive $\hat{s}_{ui}$. It follows that $\Theta$ requires a vast update within the current gradient step to learn how to correctly rank the $(u,i,j)$ triplet. 
Conversely, $\omega_{uij} \simeq 0 \implies \big( \sigma(\hat{s}_{uij}(\Theta)) \simeq 1 \iff \hat{s}_{ui}(\Theta) \gg \hat{s}_{uj}(\Theta) \big)$ is the scenario where the model does not need to update $\Theta$ on $(u,i,j)$ since it well recognized that $u$ prefers $i$ more than $j$.

\subsubsection{Adversarial Gradient Magnitude}
In the update rule of APR, each gradient step has two multiplicative scalars: the already presented \textit{Bayesian gradient magnitude} ($\omega$), and another novel scalar, named \textit{adversarial gradient magnitude} ($\omega^{adv} := (1-\sigma(\hat{s}_{uij}(\Theta + \Delta_{adv})))$. This quantity depends on how much the preference scores inferred from the perturbed model $(\Theta + \Delta_{adv})$ would be able to detect that $u$ favors $i$ more than $j$. 
It follows that, the $\omega_{uij}^{adv}$ value depends on the adversarial noise $\Delta_{adv}$ capability to revert the order preferences estimated by $\Theta$. The adversarial case in which $\Theta$ necessitates a huge update to robustify the recommender model is
\begin{align}
\begin{split}
    \small
  \omega_{uij}^{adv} \simeq 1 \implies \Big( \sigma(\hat{s}_{uij}(\Theta + \Delta_{adv})) \simeq 0 & \\ \iff \hat{s}_{ui}(\Theta + \Delta_{adv}) \ll \hat{s}_{uj}(\Theta + \Delta_{adv}) \Big)
\end{split}
\end{align}
The previous case denotes the \textit{worst-case} scenario when the model is not robust to the adversarial perturbation. In the \textit{best-case},
\begin{align}
\begin{split}
\small
  \omega_{uij}^{adv} \simeq 0 \implies \Big( \sigma(\hat{s}_{uij}(\Theta + \Delta_{adv})) \simeq 1 & \\ \iff \hat{s}_{ui}(\Theta + \Delta_{adv}) \gg \hat{s}_{uj}(\Theta + \Delta_{adv}) \Big)
\end{split}
\end{align}
the model does not require vast updates, since the original user's preferences order is preserved in spite of the perturbations. Note that both $ \omega_{uij} $ and $\omega_{uij}^{adv}$ depend on $\Theta$ and thus they change for each gradient step.

\subsubsection{Empirical Analysis of Gradient Magnitudes}~\label{subsec:grad-magnitude}
\begin{figure}[t]
    \centering
    \begin{subfigure}[b]{0.49\columnwidth}
        \centering
        \includegraphics[width=\textwidth]{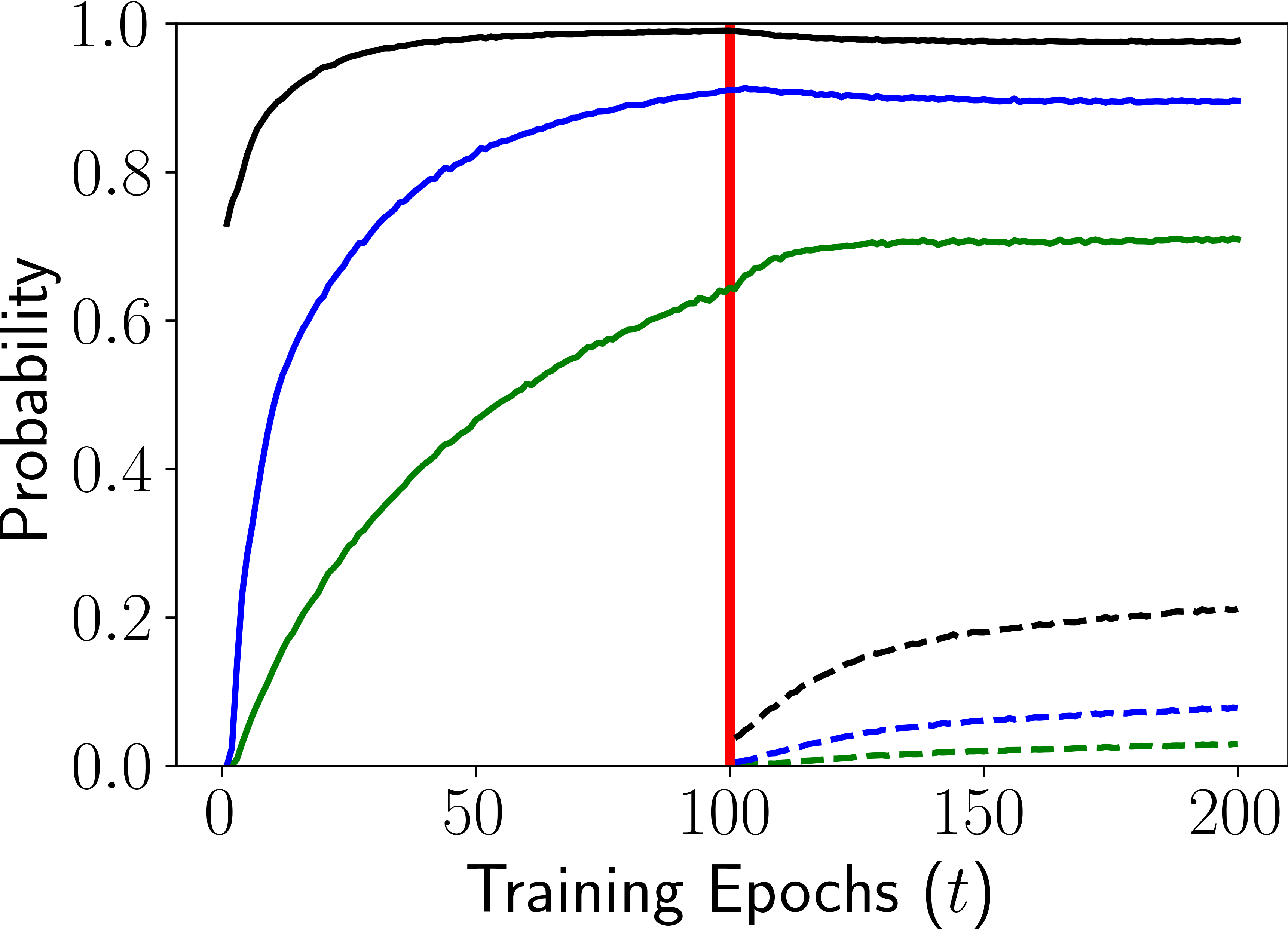}
        \caption[Amazon]%
        {{\small Amazon}}    
        \label{subfig:Amazon-bpr-apr}
    \end{subfigure}
    \begin{subfigure}[b]{0.49\columnwidth}  
        \centering 
        \includegraphics[width=\textwidth]{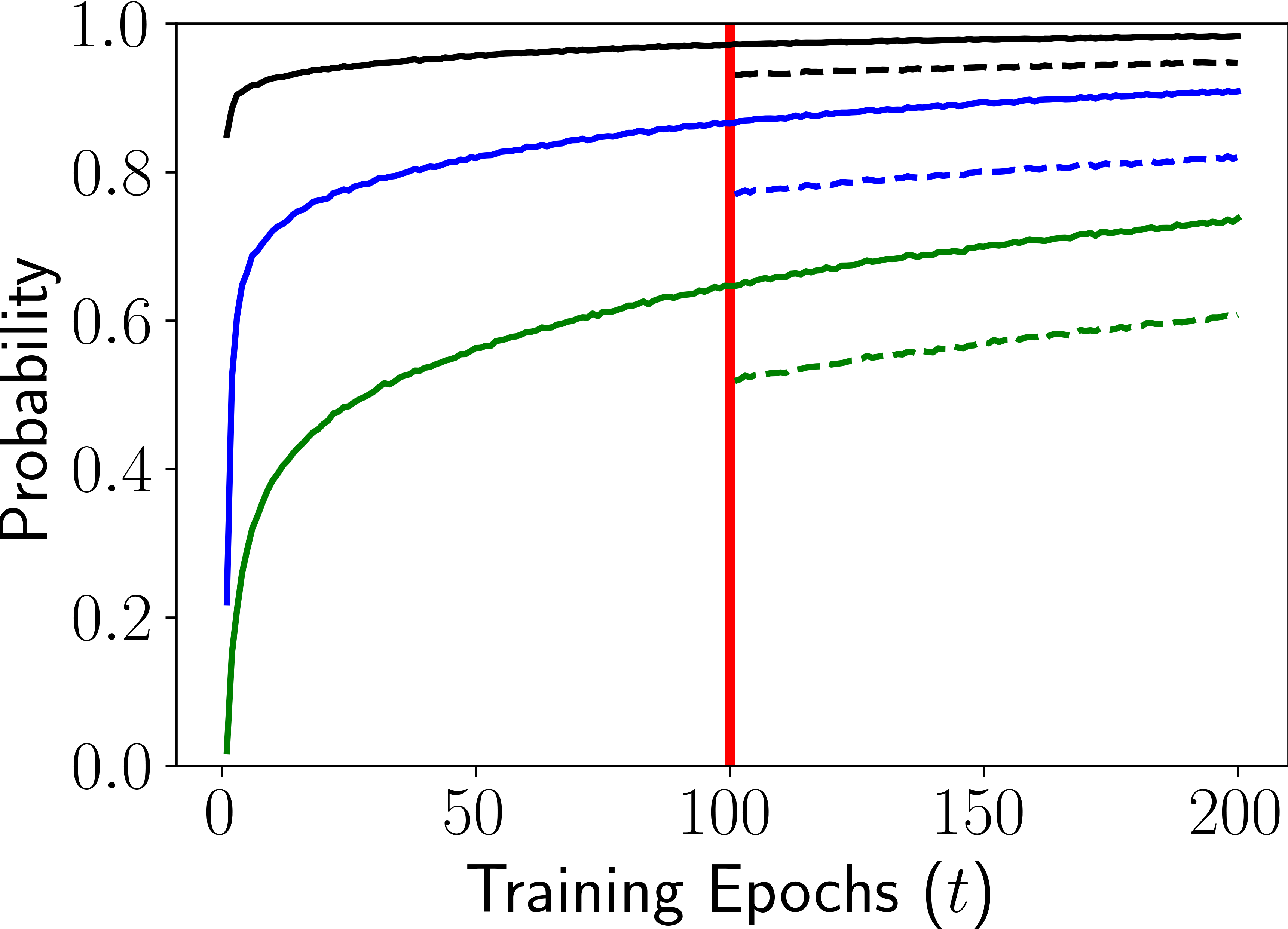}
        \caption[]%
        {{\small ML100K}}    
        \label{subfig:ML100K-bpr-apr}
    \end{subfigure}
    \hfill
    \begin{subfigure}[b]{\columnwidth}  
        \centering
        \includegraphics[width=\columnwidth]{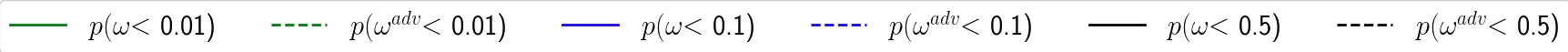}
    \end{subfigure}
      \caption{Plots on the probability that a $(u,i,j)$ triplet in $D_F$ has gradient magnitudes $\leq	\{0.01, 0.1, 0.5\}$ as in~\cite{DBLP:conf/wsdm/RendleF14}.  }
    \label{fig:gradient-magn}
    \vspace{-1em}
\end{figure}

\Cref{fig:gradient-magn} shows the probability of $\omega$ and $\omega^{adv}$ measured during the training performed on the examined datasets, i.e., Amazon~\cite{DBLP:conf/sigir/McAuleyTSH15} and+ ML100K~\cite{DBLP:journals/tiis/HarperK16}. \Cref{subfig:Amazon-bpr-apr,subfig:ML100K-bpr-apr} represent $p(\omega)$ measured for the BPR training with a number of training epochs $t \in [1,2,..., T_{BPR}]$ where $T_{BPR}=100$, and both $p(\omega)$ and $p(\omega^{adv})$ when $t \in (T_{BPR}, T_{BPR}+1,..., T_{APR}]$ with $T_{APR}=200$. The vertical red line in~\Cref{fig:gradient-magn} divides the probability measured with the initial BPR training with the ones measured when APR is activated after the $T_{BPR}$-epoch.

\Cref{subfig:Amazon-bpr-apr,subfig:ML100K-bpr-apr} show that after few training epochs $\omega$ is smaller than 0.01 for more than 85\% of the training triplets of the Amazon dataset, and 65\% for the ML100K ones. Next, the magnitudes measured on all the triplets are smaller than 0.5, i.e., $p(\omega_{uij} < 0.5) \simeq 1.0, \forall (u,i,j) \in D_f$, after the first 50 epochs for both the datasets. $\omega \simeq 0$ after the first few training epochs is a BPR \textit{gradient vanishing} issue leading to the slow convergence~\cite{DBLP:conf/wsdm/RendleF14}.

Analyzing $\omega^{adv}$ in~\Cref{subfig:Amazon-bpr-apr,subfig:ML100K-bpr-apr}, APR is not affected by the BPR gradient vanishing issues. For ML100K, all the APR lines (dotted curves) are lower than the BPR ones (continue curves), meaning that APR magnitudes are consistently higher than the BPR ones. This phenomenon is evident in the experiments on the Amazon dataset. Indeed,~\Cref{subfig:Amazon-bpr-apr} shows that the probability of getting small magnitudes, i.e., $p(\omega^{adv} < 0.1)$, is smaller than 10\% also when 100 APR-training epochs have been performed on the model. 
We explain this behavior with the fact that the APR objective function also considers the adversarial regularizer that forces the $\Theta$ updates to limit the performance drop in adversarial settings. These results confirm that APR is a solution to both robustify and stabilize the BPR model training, as claimed, for example, in~\cite{DBLP:conf/sigir/0001HDC18, DBLP:conf/sigir/YuanYB19, DBLP:journals/tkde/TangDHYTC20, DBLP:conf/recsys/ChenL19}. 

\subsection{Amplification of Popularity Bias}~\label{subsec:amplification-bias}
The stability and accuracy effects of the \textbf{adversarial training} have not been explored on \textbf{beyond-accuracy} results. For instance, data-tailed distribution is a property that received strong attention in the literature of RSs. Indeed, it is common in RSs that few items, named \textit{short-head} items ($I_{SH}$), receive much more feedbacks than many other ones, named \textit{long-tail} ($I_{LT}$)~\cite{DBLP:conf/recsys/AbdollahpouriBM17, DBLP:journals/ipm/BorattoFM21}. In this work, we use the short-head and long-tail definition used by~\citet{DBLP:conf/recsys/AbdollahpouriBM17}, where the short-head set, composed of the top 20\% of items by popularity, has much more feedback than the long-tail one, which contains the remaining 80\% of items. Since BPR is known to be affected by the amplification of biases~\cite{DBLP:journals/umuai/JannachLKJ15,DBLP:conf/sigir/ZhuWC20,DBLP:journals/ipm/BorattoFM21}, we conjecture that APR could be affected, or even intensify, biases amplification since it hugely influences the BPR-based pre-trained model, as empirically verified before (see~\Cref{subsec:grad-magnitude}). 

\subsubsection{Effects of Imbalanced Data}\label{subsubsec:imbalance}
Since the users' feedback data distribution is affected by popularity bias, the sampling distribution of positive items is $p(i \in I_{SH}|u) \geq p(i \in I_{LT}|u)$. It means that the probability that a positive item of one triplet in $D_F$ is in the set of short-head items is higher than the probability of being in the long-tail. Then, the uniform sampling strategy of negative items used in BPR and APR results follows $p(j \in I_{SH}|u) = p(j \in I_{LT}|u) = \frac{1}{|I|}$. It means that the probability that the negative item in the $(u, i, j)$-training triple does not depend on the feedback distributions since they are randomly extracted from the full set of items, i.e., $I$. The previous relations evidence that the difference between the sampling distributions to build $D_F$ could influence both the number and the sign of the model parameter updates.
To study whether APR amplifies the popularity bias, we define \textbf{global positive} and \textbf{global negative} updates.

\begin{definition}[\textbf{Global Positive Update} ($\Omega^{+}$)]
Let $t \in \{1, 2, ., T_{BPR}, T_{BPR}+1, .., T_{APR}\}$ be a training epoch and $D_F(t)$ be the set of training triplets built for the $t$-th epoch, then the \textit{global positive update} on short-head items is
\begin{equation}
  \Omega^{+}(I_{SH}|D_F(t)) := \sum_{(u,i,j) \in D_F(t) \land i \in I_{SH}} \omega_{uij}(t) + \omega_{uij}^{adv}(t) 
\end{equation}
while the global positive update for long-tail items is 
\begin{equation}
  \Omega^{+}(I_{LT}|D_F(t)) = \sum_{(u,i,j) \in D_F(t) \land i \in I_{LT}} \omega_{uij}(t) + \omega_{uij}^{adv}(t)
\end{equation}
\end{definition}

\begin{definition}[\textbf{Global Negative Update} ($\Omega^{-}$)]
The global negative update for short-head items at $t$-th training epoch is
\begin{equation}
  \Omega^{-}(I_{SH}|D_F(t)) := - \sum_{(u,i,j) \in D_F(t) \land j \in I_{SH}} \omega_{uij}(t) + \omega_{uij}^{adv}(t)
\end{equation}
while the global negative update for long-tail ones is 
\begin{equation}
  \Omega^{-}(I_{LT}|D_F(t)) := - \sum_{(u,i,j) \in D_F(t) \land j \in I_{LT}} \omega_{uij}(t) + \omega_{uij}^{adv}(t)
\end{equation}
\end{definition}
where, negative means that the model will learn to 'separate' mode the current negative item and the positive.
Since $\Omega^{+}$ focuses on positive items ($i$) and $\Omega^{-}$ focuses on negative ones ($j$), we expect that the \textbf{global number of positive updates on short-head items is higher than the one on long-tail ones}. It means that APR could be algorithmically affected by the \textbf{amplification of the popularity bias} as already checked on BPR. Below, we empirically verify whether APR amplifies BPR issues.


\begin{figure}[!t]
  \centering
      \begin{subfigure}[b]{0.49\columnwidth}
        \centering
        \includegraphics[width=\columnwidth]{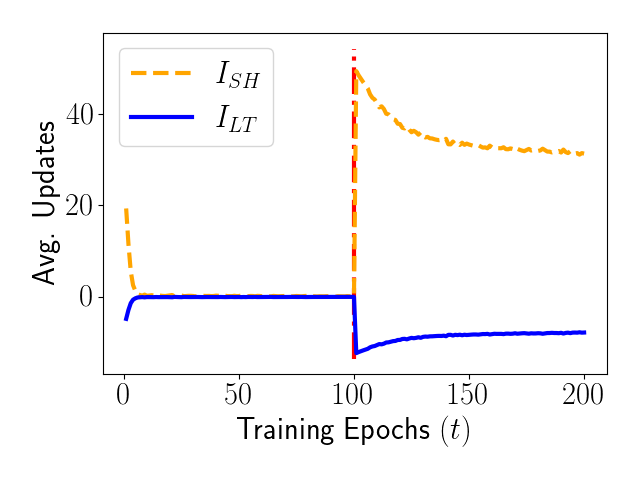}
        \caption[Amazon]
        {{\small Amazon}}    
        \label{subfig:sum-updates-amazon}
    \end{subfigure}
    \begin{subfigure}[b]{0.49\columnwidth}  
        \centering 
        \includegraphics[width=\columnwidth]{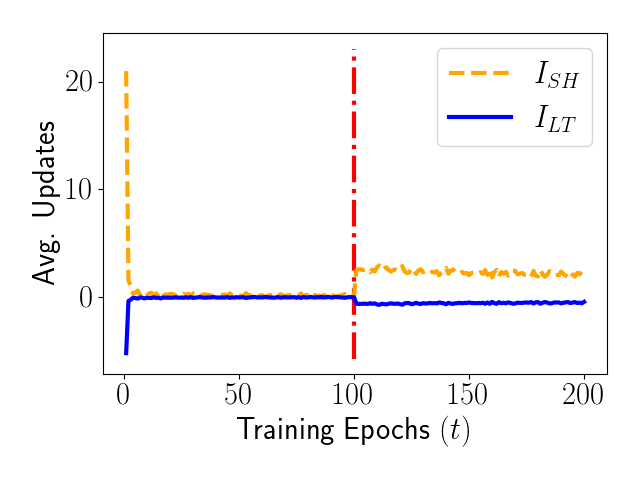}
        \caption[ML100K.]%
        {{\small ML100K.}}
        \label{subfig:sum-updates-ml100k}
    \end{subfigure}
  \caption{\small Plots of the global gradient updates averaged by the number of items in $I_{SH}$ and $I_{LT}$. Red line indicates the start of APR.}
    \label{fig:sum-updates}
    \vspace{-2em}
\end{figure}

\subsubsection{Empirical Validation: the Wine-Glass Phenomenon}
\Cref{subfig:sum-updates-amazon,subfig:sum-updates-ml100k} show
the $\Omega^{+}(I_{SH}|D_F(t)) + \Omega^{-}(I_{SH}|D_F(t))$ and $\Omega^{+}(I_{LT}|D_F(t))$ $+ \Omega^{-}(I_{LT}|D_F(t))$ averaged by number of items in $I_{SH}$ and $I_{LT}$, respectively.
We firstly observe that the sum of the first quantity is always positive for short-head items, while the second is negative for long-tail ones. Then, we identify a \textbf{wine-glass phenomenon} in~\Cref{subfig:sum-updates-amazon,subfig:sum-updates-ml100k}. In fact, each plot can be divided into three parts: a \textbf{base}, a \textbf{stem}, and a \textbf{bowl}.
The \textbf{base} represents the BPR training epochs in which the updates on $I_{SH}$ and $I_{LT}$ have an absolute magnitude different from 0. Already in this training phase, it can be seen that the average gradient magnitudes associated with $I_{SH}$ are more significant than the one on $I_{LT}$,  consistently with the results in~\cite{DBLP:journals/umuai/JannachLKJ15, DBLP:conf/sigir/ZhuWC20, DBLP:conf/cikm/MansouryAPMB20, DBLP:journals/ipm/BorattoFM21}.
The \textbf{stem}, the second component, characterizes the last epochs of BPR ($T_{BPR}/2 < t \leq T_{BPR}$) showing the gradient vanishing problem as examined in~\Cref{subsec:grad-magnitude}. There is no amplification of the bias in this phase, since the model performs very tiny gradient updates.
The last part of the glass, the \textbf{bowl}, exposes the average magnitudes in the case of APR training ($t> T_{BPR}$). Here, the average sum of Bayesian and adversarial gradient magnitudes on each item in $I_{SH}$ is much more notable than the one for $I_{LT}$. These results empirically confirm that APR could increase even more than BPR the item popularity bias. In the next section, we examine beyond-accuracy and bias.
\section{ Experiments}\label{sec:experiments}

\noindent \textbf{Datasets.}  
We experiment on \textbf{ML100K}~\cite{DBLP:journals/tiis/HarperK16}, a popular dataset used for recommender model prototyping, and \textbf{Amazon}~\cite{DBLP:conf/sigir/McAuleyTSH15}, an e-commerce dataset holding customers ratings. For each dataset, we employ the \textit{temporal leave-one-out} protocol~\cite{DBLP:conf/sigir/0001HDC18}. The first dataset has 943 users, 1,682 items, and 100K ratings with $p(i|I_{SH})=0.6452 > p(i|I_{LT})=0.3548$, the second contains 3,915 users, 2,549 items, and 77,328 ratings with $p(i|I_{SH})= 0.5747 > p(i|I_{LT})=0.4253$.

\noindent \textbf{Evaluation Metrics.}  We study top-$k$ performance, reporting: the precision (\textit{Prec@k}), recall (\textit{Rec@k}), and normalized discounted cumulative gain (\textit{nDCG@k})~\cite{DBLP:reference/sp/NingDK15} for the accuracy, the item coverage ($Cov_{\%}@k$) and the novelty ($Nov$)~\cite{Zhou4511}, for the beyond-accuracy, and ARP, ACLT, and APLT for the long-tail diversity measures for the popularity bias~\cite{DBLP:conf/flairs/AbdollahpouriBM19}.

\noindent \textbf{The Algorithm.} 
We test BPR and APR on matrix-factorization (MF)~\cite{DBLP:journals/computer/KorenBV09}. MF is an LFM representing items and users by rows of low-rank matrices of embeddings ($\Theta$).

\noindent \textbf{Reproducibility} 
We set $f=64$ as in~\cite{DBLP:conf/sigir/0001HDC18}. We train BPR-MF for $T_{BPR}$ epochs grid-searching the learning rate $\eta \in \{0.005, 0.01, 0.05 \}$ validating on $Rec@50$.Then, we explore $\epsilon \in \{0.001, 0.01, 0.1, 1.0\}$ and $\alpha \in \{0.001, 0.01, 0.1, 1.0, 10.0\}$ to train APR-MF. We use $T_{BPR}=100$ and $T_{APR}=200$. Further reproducibility details, the code, and the data will be available on the \href{https://github.com/sisinflab/Biased-AdvReg}{GitHub} repository.

\vspace{-1em}
\subsection{Results and Discussion}

\begin{table}[!t]
\caption{Results on top-$50$ lists. The $\uparrow$ means that bigger value is related to an higher popularity bias, $\downarrow$ means less bias.}
\begin{center}
\begin{adjustbox}{width=1\columnwidth,center}
\begin{tabular}{|l|rrr|rr|rrr|}
\hline

\multirow{2}{*}{\textbf{Model}}  & \multicolumn{3}{c|}{\multirow{1}{*}{\textbf{Accuracy}}} & \multicolumn{2}{c|}{\multirow{1}{*}{\textbf{Beyond}}} & \multicolumn{ 3}{c|}{\textbf{Popularity Bias}} \\ \cline{ 2- 9}

& \multicolumn{1}{c}{\textit{Rec}} & \multicolumn{1}{c}{\textit{Prec}} & \multicolumn{1}{c|}{\textit{nDCG}} & \multicolumn{1}{c}{\textit{Nov}} & \multicolumn{1}{c|}{\textit{$Cov_{\%}$}} &      \multicolumn{1}{c}{\textit{ARP $\uparrow$}} & 
     \multicolumn{1}{c}{\textit{APLT} $\downarrow$} & 
     \multicolumn{1}{c|}{\textit{ACLT} $\downarrow$}\\ \hline

\multicolumn{9}{|c|}{\textbf{ML100K}}\\\hline

BPR-MF & 0.3871 & 0.0077  & 0.1222 & 2.7653 & 71.22 & 176.64 & 0.2890 & 14.4486\\ \hline
APR-MF& 0.3966  & 0.0079 & 0.1260* & 2.7577* & 71.22* & 177.33* & 0.2841* & 14.2068*\\ \hline
\multicolumn{1}{|r|}{\textbf{R.V.}} & \textbf{+2.47\%} & \textbf{+2.47\%} & \textbf{+3.15\%} & \textbf{-0.27\%} & \textbf{0.00\%} & \textbf{+0.39\%} & \textbf{-1.67\%} & \textbf{-1.67\%}\\ \hline

\multicolumn{9}{|c|}{\textbf{Amazon}}\\\hline

BPR-MF & 0.2077  & 0.0042 & 0.0656 & 6.0431 & 99.37 & 106.59 & 0.3541 & 17.7055 \\ \hline
APR-MF & 0.2130  & 0.0043 & 0.0687* & 5.6805* & 90.58* & 131.30* & 0.2829* & 14.1471* \\ \hline
\multicolumn{1}{|r|}{\textbf{R.V.}} & \textbf{+2.58\%} & \textbf{+2.58\%} & \textbf{+4.63\%} & \textbf{-6.00\%} & \textbf{-8.85\%} & \textbf{+23.18\%} & \textbf{-20.10\%} & \textbf{-20.10\%}\\ \hline

\multicolumn{9}{l}{* statistically significant results ($\text{p-value} \leq 0.05$) using the paired-t-test.}

\end{tabular}
\end{adjustbox}
\end{center}
\label{table:accuracy-vs-beyond-vs-bias}
\vspace{-2em}
\end{table}

\noindent \textbf{Accuracy and Beyond-Accuracy Results} Analyzing~\Cref{table:accuracy-vs-beyond-vs-bias}, we identify that APR tends to reduce the novelty and coverage values compared to the one measured on BPR. For ML100K, APR-MF improves $Rec$, $Prec$, and $nDCG$ by more than $2\%$, with a slight reduction of $Nov$, i.e., $\textbf{R.V.}(Nov)=-0.27\%$. Consistently, we measure an $\textbf{R.V.}(Rec)=+2.58\%$ and $\textbf{R.V.}(Cov_{\%})=-8.85\%$ for Amazon. \noindent \textbf{\underline{Observation 1:}} \textit{APR can negatively influence the beyond-accuracy recommendation performance}.

\noindent \textbf{Popularity Bias Results}
As expected by the analysis in~\Cref{subsec:amplification-bias}, the three long-tail diversity scores get negative \textbf{R.V.} when comparing APR-MF with BPR-MF, its building block. 
Examining the $ARP$ values, we identify that APR-MF results increase the recurrence of most popular items in the recommendation lists. For instance, the $\textbf{R.V.}(ARP)=+23.18\%$ on Amazon and +0.39\% on ML100K.  As stated by~\citet{DBLP:conf/flairs/AbdollahpouriBM19}, since the ARP is not a good measure of long-tail diversity when used only on its own, we also report $APLT$ and $ACLT$. For both metrics, the \textbf{R.V.} values are negatives, a behavior consistent with the growth of $ARP$. 

\noindent \textbf{\underline{Observation 2:}} \textit{APR can amplify the popularity bias more than BPR}. 

\noindent \textbf{Impact of Data Characteristics}
Trying to connect the results observed in~\Cref{table:accuracy-vs-beyond-vs-bias} together with the dataset characteristics, we observe that Amazon, the dataset with the lowest density ($0.0077$), got the worst amplification of bias and reduction of beyond-accuracy when used by APR.

\noindent \textbf{\underline{Observation 3:}} \textit{The dataset characteristics could impact biases amplification and beyond-accuracy goodness in APR settings}.
\section{Conclusion and Open Directions}\label{sec:conclusion}
Adversarial personalized ranking (APR) is a popular optimization framework to robustify and stabilize model-based recommenders against adversarial perturbations. In this work, we modeled the learning characteristics of APR, identifying a potential phenomenon of amplification of popularity bias and reduction of beyond-accuracy performance, justified by the identification of a learning pattern, named \textit{wine-glass phenomenon}, that confirms the APR trend to perform more positive gradient updates on short-head items than long-tail ones, with a difference in magnitude greater than the one measured on BPR (the APR building-block). Experiments on MF recommenders trained on two datasets confirmed the theoretical findings by measuring both beyond-accuracy and popularity bias performance worsening in APR settings. Recognizing the importance and popularity of APR, we consider it important to solve the limits of APR and investigate novel robustification strategies.

\bibliographystyle{ACM-Reference-Format}
\bibliography{bibliography}

\end{document}